\documentclass[11pt]{article}
\usepackage[affil-it]{authblk}
\usepackage{amsfonts}
\usepackage{amsmath}
\usepackage{hyperref}
\usepackage{fullpage}
\usepackage{stackrel}
\usepackage{amssymb}
\usepackage{graphicx}
\usepackage{tikz}
\usepackage{pgfplots}
\usetikzlibrary{matrix,chains,positioning,decorations.pathreplacing,arrows}
\usepackage{qcircuit}
\usepackage{float}
\providecommand{\U}[1]{\protect\rule{.1in}{.1in}}

\newtheorem{theorem}{Theorem}

\newtheorem{example}[theorem]{Example}

\newtheorem{remark}[theorem]{Remark}

\usepackage{color}

\providecommand{\xv}{\boldsymbol{x}}
\providecommand{\xiv}{\boldsymbol{\xi}}
\providecommand{\wv}{\boldsymbol{w}}
\providecommand{\de}{\mathrm{d}}
\providecommand{\rme}{\mathrm{e}}
\providecommand{\rmi}{\mathrm{i}}

\providecommand{\RS}{\mathrm{RS}}
\providecommand{\CX}{\mathrm{CX}}

\begin{document}

\title{Pattern capacity of a single quantum perceptron}
\author[1,2]{Fabio Benatti}
\author[1,2]{Giovanni Gramegna}
\author[3,4]{Stefano Mancini}
\affil[1]{Dipartimento di Fisica, Universit\'{a} di Trieste, Strada Costiera 11, I-34151, Trieste, Italy}
\affil[2]{Istituto Nazionale di Fisica Nucleare, Sezione di Trieste, Strada Costiera 11, I-34151, Trieste, Italy}
\affil[3]{School of Science and Technology, University of Camerino, I-62032 Camerino, Italy}
\affil[4]{Istituto Nazionale di Fisica Nucleare, Sezione di Perugia, Via A. Pascoli, I-06123 Perugia, Italy}
\date{}
\maketitle
\begin{abstract}
	Recent developments in Quantum Machine Learning have seen the introduction of several models to generalize the classical perceptron to the quantum regime. The capabilities of these quantum models need to be determined precisely in order to establish if a quantum advantage is achievable. Here we use a statistical physics approach to compute the pattern capacity of a particular model of quantum perceptron realized by means of a continuous variable quantum system.
\end{abstract}
\section{Introduction}
Artificial neural networks have proven to be an extremely efficient computational model in specific tasks such as pattern recognition or image classification and have revolutionized the field of data analysis on classical computers~\cite{Hertz,NNreview,NNreviewIEEE,DeepLearning}. At the same time, the advent  of quantum computation has shown that purely quantum mechanical features such as coherence and entanglement allow to address hard computational tasks with an exponential improvement of the performances compared to classical computation~\cite{Nielsen}. The great success achieved in these two fields has motivated a surge of interest in quantum machine learning, exploring the interaction between machine learning and quantum computation, with the aim to understand whether the two fields can benefit from each other.

The simplest model of an artificial neuron traces back to the classical Rosenblatt’s perceptron~\cite{Rosenblatt57}, which can be seen as the simplest learning algorithm for binary classification. Several possiblities can be considered to implement a perceptron  
{  by means of} a quantum architecture~\cite{Lewenstein,Kak95,Zak98,Wan17,BenattiManciniMangini,Machiavello19,Artiaco21}. {  In this context it} is important to investigate the capability of a particular model of quantum perceptron 
{  to achieve quantum advantages with respect  to its classical counterparts}.

The {  main limitation affecting a} single classical perceptron 
{  is due to} the fact that the classification {  task} is performed through a separation of patterns belonging to different classes through a hyperplane in the vector space containing the $N$ features defining the pattern. In particular, it was soon pointed out that a simple perceptron is unable to compute the XOR function~\cite{MinskyPapert}, since this corresponds to a classification problem where different classes cannot be separated with a line in the plane. However, it was found that when a large number of features is considered, i.e. for patterns in a vector space with a large dimension $N$,  given $p$ random labeled patterns it is extremely unlikely that a perceptron cannot classify them if $p<2 N$ for large $N$~\cite{Cover,Venkatesh92}. On the contrary, the probability that $p$ random labeled patterns can be classified by a simple perceptron becomes {  vanishingly small} for $p>2N$ in the large $N$ regime. It became clear that the important parameter to characterize the performances of a perceptron is then the ratio $\alpha=p/N$, and led to identify the critical value of this ratio as the pattern capacity of a classical perceptron  which thus is  $\alpha_c=2$.

In the seminal work~\cite{Gardner88a}, Gardner took a new approach 
to the pattern capacity of neural networks, adopting tools of statistical physics and in particular methods from the theory of disordered systems. The possibility to find an hyperplane separating randomly labeled patterns belongs  in fact to the class of random constraint satisfaction problems~\cite{Artiaco21,Franz16,Franz17}, which can be investigated using the statistical theory of spin-glasses. In this approach, the parameter $\alpha$ gives rise to a phase transition in the high-dimensional case, and the pattern capacity is determined by the critical value $\alpha_c$ separating the SAT-phase, for $\alpha<\alpha_c$, where it is possible to satisfy all the constraints, i.e. classify all the patterns, from the UNSAT-phase, $\alpha>\alpha_c$, where the minimum number of unsatisfied constraints is larger than zero.

Here, we will follow Gardner's statistical approach to derive the pattern capacity of a particular model of quantum perceptron, based on a continuous variable multi-mode quantum system, which was introduced in~\cite{BenattiManciniMangini}. We show that this model offers no quantum advantage over its classical counterpart,  since its capacity is always smaller than that of its classical limit.

The article is structured as follows. In section~\ref{sec:CP} we introduce the classical perceptron and the definition of its pattern capacity. In section~\ref{sec:QP} we describe the model of quantum perceptron under investigation, and present the resulting pattern capacity. In section~\ref{sec:Methods} we explain in detail the techniques employed, based on the same statistical approach used by Gardner to determine the pattern capacity of a classical perceptron. Finally, in section~\ref{sec:Conclusion} we discuss the results 
herewith achieved comparing them with the pattern capacity also obtained by means of the statistical approach, but for a different model of quantum perceptron.

\section{The classical perceptron and its capacity}
\label{sec:CP}

\begin{figure}[ht]
\centering
\includegraphics[width=.8\textwidth]{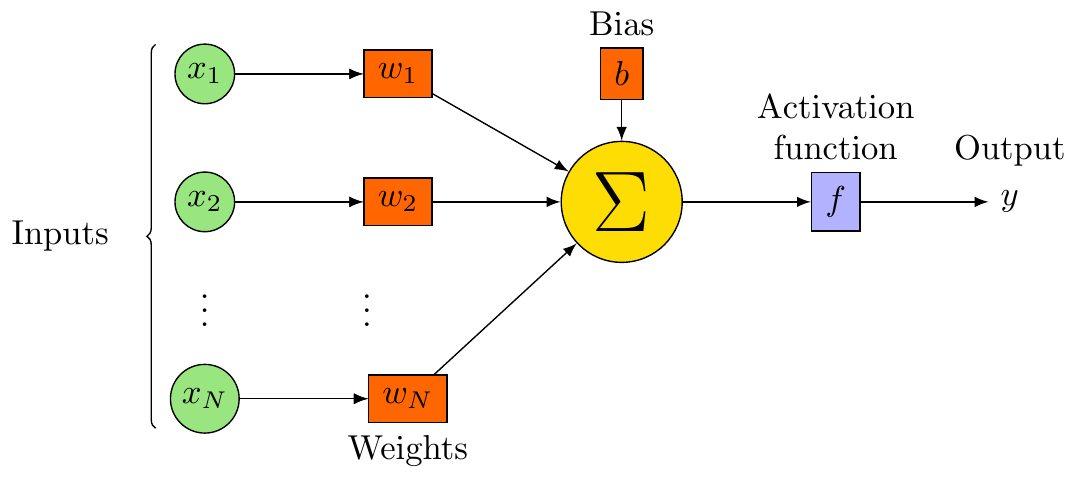}
\caption
{Schematic representation of the mathematical model of a perceptron.}
\label{fig:McCullochNeuron}
\end{figure}

A classical perceptron is a mathematical model that mimics the functioning of a physical neuron. 
By referring to \figurename{~\ref{fig:McCullochNeuron}}, given an input pattern, namely an $N$-dimensional vector  $\boldsymbol{x} = (x_1, \hdots, x_n)$ with $x_i$ real or discrete (eventually binary), 
a perceptron computes an affine transformation 
\begin{equation}
\label{affunct}
\boldsymbol{x}\longmapsto z:=\boldsymbol{x}\cdot \boldsymbol{w} + b,
\end{equation}
with real parameters $\boldsymbol{w}=(w_1, \hdots, w_n)$ and $b$, called \textit{weights} and \textit{bias}, respectively.
Subsequently, the perceptron evaluates on the output  $z$ an \textit{activation function} $f: \mathbb{R} \rightarrow \mathbb{R}$, eventually yielding 
the final result $y:=f(z)$.

There exist different possible activation functions, some being more computationally efficient like $\mathrm{sgn}(z)$, and others more biologically inspired as the hyperbolic tangent $f(z)= ({\rm e}^{z}-{\rm e}^{-z})/({\rm e}^{z}+{\rm e}^{-x})$, or the sigmoid function $f(z)=1/(1+{\rm e}^{-z})$.  

In the following, we consider the activation function 
$\mathrm{sgn}(z)$ and, for the sake of simplicity, $b=0$ (note however that one can always include a bias by adding a component to patterns and weights, and consider for the additional component $x_0=1$ and $w_0=b$). 
In such a context, a binary classification problem amounts to 
assigning an input pattern to one of two possible classes, indexed by a dichotomic variable $\xi\in\{-1,1\}$. In the following, we focus on input patterns with binary entries, 
\begin{equation}
	\label{patterns}
	\Pi:=\{\xv^\mu\in\{-1,1\}^N\,:\, 1\leq \mu\leq p\}\ {  ,}
\end{equation}
and denote with
\begin{equation}
\label{dichotomies}
\Xi:=\left\{\boldsymbol{\xi}=\{\xi^\mu\}_{\mu=1}^p\,:\,\xi^\mu=\pm\right\}
\end{equation}
the set containing the $2^p=\hbox{card}(\Xi)$ possible classifications of the given $p$ patterns.
{ Such a classical perceptron} classifies the patterns by means of a suitably chosen weight vector $\boldsymbol{w}\in \mathbb{R}^N$ which performs a linear separation of the input space according to the relation:
\begin{equation}
	\sigma^\mu=\mathrm{sgn}(\boldsymbol{w}\cdot \xv^\mu), \qquad 1\leq \mu \leq p.
\end{equation}
If a particular target classification $\xiv\in\Xi$ is fixed, we say that the perceptron correctly classifies the input patterns if $\sigma^\mu=\xi^\mu$ for each $\mu=1,\dots,p$. For later convenience it is useful to express this requirement in a different way introducing the pattern stabilities
\begin{equation}\label{def:PatStab}
		\Delta^\mu=\xi^\mu\frac{\wv\cdot \xv^\mu}{\|\wv\|},
\end{equation}
where $\|\wv\|=\sqrt{\sum_j w_j^2}$ denotes the { weight} euclidean norm. The {classification of} the input patterns is correct with respect to a chosen target classification $\boldsymbol{\xi}$ if 
\begin{equation}\label{eq:CondStab}
	\Delta^\mu\geqslant 0 \qquad \text{for each } \mu=1,\dots,p.
\end{equation}
From a geometric point of view, equation \eqref{eq:CondStab} expresses the fact that the hyperplane in $\mathbb{R}^N$ perpendicular to $\wv$ separates the patterns $\xv^\mu$ with $\xi^\mu=+1$ from patterns with $\xi^\mu=-1$. 

\begin{example}
\label{ex:XOR}
The four binary patterns 
$$
\Pi=\{\xv^1=(-1,-1)\ ,\ \xv^2=(-1,1)\ ,\ \xv^3=(1,-1)\ ,\ \xv^4=(1,1)\}
$$ 
are associated with $2^4=16$ dichotomies. Sending
$x\mapsto (1+x)/2$, the chosen patterns constitute a two-bit random variable with values $(i,j)$, $i,j=0,1$, and the choice $\boldsymbol{\xi}=(-1,1,1,-1)$ 
implements the $\mathrm{XOR}$ binary gate
$$
\mathrm{XOR}(0,0)=0\ ,
\ \mathrm{XOR}(0,1)=1\ ,\ \mathrm{XOR(}1,0)=1\ ,\ \mathrm{XOR}(1,1)=0\  .
$$
The so-called $\mathrm{XOR}$ problem~\cite{MinskyPapert} in machine-learning exactly corresponds to the impossibility of computing the function $\mathrm{XOR}$ by means of  a perceptron, since it is not possible to find a line in $\mathbb{R}^2$ which separates $\xv^1$ and $\xv^4$ from $\xv^2$ and $\xv^3$.
\end{example}

The XOR problem indicates that it is not always possible to associate to a set of patterns $\Pi$ the correct target classification $\xiv$ by using a perceptron. 
The existence of a particular choice of weights which realizes the desired classification depends on the particular set $\Pi$ and on the target classification $\xiv$ considered. A natural way to proceed is to consider a statistical approach, where one fixes the total number of patterns $p$ and constructs the pattern set $\Pi$ by selecting randomly $p$ points from $\{-1,1\}^N$, and choose independently the corresponding target classifications $\xi^\mu$. 

In the approach introduced by Gardner in~\cite{Gardner88a}, the capacity of a perceptron is computed through the evaluation of the fraction of volume in the space of the weights $\wv\in\mathbb{R}^N$ which correctly classifies the given input patterns $\Pi$. In this context the condition \eqref{eq:CondStab} is strenghtened in order to include the possibility of the additional requirement of \textit{finite stability} $\kappa>0$, which amounts to requiring that the quantity defined in~\eqref{def:PatStab}
\begin{equation}\label{eq:CondStabBis}
	\Delta^\mu>\kappa \quad \text{for each}\quad \mu=1,\dots,p.
\end{equation}  

\begin{remark}
	From a practical perspective, the introduction of a finite stability requirement allows one to obtain a perceptron which avoids classification errors even in the presence of errors in the input patterns. For example, it can happen that instead of the correct pattern $\xv^\mu$, a noisy version of it is presented to the perceptron, say $\tilde{\xv}^\mu$, where some sign is changed, i.e. $\tilde{x}^\mu_j=-x^\mu_j$ for some $j=1,\dots,N$. In such  a case, a sharp classification request, $\kappa=0$, would lead to an error.
\end{remark}

Due to the definition of the quantities $\Delta^\mu$, the validity of the conditions \eqref{eq:CondStab} is independent on the normalization of the weight vector $\wv$, which implies that we can {  suitably} fix it
and compute the relative volume of weights which satisfy the conditions \eqref{eq:CondStabBis} among the weights with a fixed norm. We choose it to be $\|\wv\|^2=N$ (which corresponds to taking vectors $\wv$ whose components are $O(1)$), so that the relative volume of interest can be written as:
\begin{equation}\label{eq:relV}
	V:=V\left(\{\xi^\mu\xv^\mu\}_{\mu=1}^p\right)=\frac{1}{C_N} \int_{\mathbb{R}^N} \de \wv~ \delta(\|\wv\|^2-N) \prod_{\mu=1}^p\theta \left(\Delta^\mu-\kappa\right)
\end{equation}
where $\theta$ denotes the Heaviside step function, $\theta(z)=1$ for $z\geqslant 0$ and zero otherwise,
\begin{equation}
	C_N=\int_{\mathbb{R}^N}\de \wv~ \delta( \|\wv\|^2-N)=\frac{\pi^{N/2}N^{N/2-1}}{\Gamma(N/2)}\sim { \sqrt{\frac{(2\pi \rme)^N}{4\pi N}}}
\end{equation}
is the the reference volume of weights contained in the hypersphere of radius $\sqrt{N}$, $\Gamma(z)$ is the Gamma function and{ the large $N$ behaviour follows from the Stirling approximation.}
The expression \eqref{eq:relV} resembles a statistical-mechanical partition function, where the conventional exponential weight is replaced by an all-or-nothing weight given by the step functions.  In analogy with statistical mechanics, the relevant quantity is actually $\ln V$ rather than $V$ itself, of which we will compute the average $\langle \ln V \rangle$ with respect to the identically and independently distributed stochastic variables $\xi^\mu$, $x^\mu_j$, with $\mu=1,2,\dots,p$, and $j=1,2,\ldots,N$.

Using the replica method~\cite{Gardner88a}, Gardner showed the existence of a critical value of $\alpha$, given by 
\begin{equation}
	\label{eq:alphacrit}
	\alpha_c(\kappa)=\left[\int_{-\kappa}^{\infty}\frac{\de t}{\sqrt{2\pi}}\rme^{-t^2/2}(t+\kappa)^2 \right]^{-1}.
\end{equation}
such that for $\alpha<\alpha_c(\kappa)$ the following limit holds:
\begin{equation}\label{eq:aveLogV1}
	\lim_{\stackrel{N,p\rightarrow \infty}{{  p/N=\alpha}}}\frac{\langle \ln V \rangle}{N} = \mathcal{F}(\alpha,\kappa)\equiv \min_{0\leqslant q \leqslant 1}\left\{\alpha \int \frac{\de t}{\sqrt{2\pi}}\rme^{-\frac{t^2}{2}}\ln\left[1-\Phi\left(\frac{t\sqrt{q}+\kappa}{\sqrt{1-q}}\right) \right]+\frac{q}{2(1-q)}+\frac{\ln(1-q)}{2} \right\},
\end{equation}
where
\begin{equation}
	\Phi(x):=\frac{1}{\sqrt{2\pi}}\int_{-\infty}^{x} \rme^{-t^2/2}\de t,
\end{equation}
while for $\alpha> \alpha_c(\kappa)$:
\begin{equation}\label{eq:aveLogV2}
	\lim_{\stackrel{N,p\rightarrow \infty}{{  p/N=\alpha}}}\frac{\langle \ln V \rangle}{N} = - \infty.
\end{equation}

Gardner's statistical approach was later reformulated in a mathematically rigorous setting~\cite{Talagrand1999,Shcherbina2003}, where it was pointed out that the random variable $N^{-1}\ln V$ is self-averaging. 
{ Namely,} deviations from the average {$N^{-1}\langle \ln V\rangle$} become vanishingly small in the limit $N\rightarrow\infty$. In other words,
{ the average $N^{-1}\langle \ln V\rangle$ is} a good representative of what happens for almost all realizations of the random variable 
{$N^{-1}{\ln V}$, each realization corresponding} to a particular choice of the random patterns and classifications $\{\xv^{\mu},\xi^\mu\}_{\mu=1}^p$. 

{According to such an observation,} equations \eqref{eq:aveLogV1} and $\eqref{eq:aveLogV2}$ 
{can be interpreted  as follows: for} $\alpha<\alpha_c(\kappa)$ the relative volume of weights which are able to {correctly} classify a random choice of the patterns is approximately $V\sim\exp(-{  \mathcal{F}(\alpha,\kappa) }\,N)$, while above the critical value, {that is when} $\alpha>\alpha_c(\kappa)$, 
{ the relative volume is more than exponentially vanishing with $N$, i.e.} $V=o(\exp(-cN))$. 

\section{Continuous variable quantum perceptron: pattern capacity}
\label{sec:QP}
\begin{figure}[]
	\centering
	\includegraphics[width=.8\textwidth]{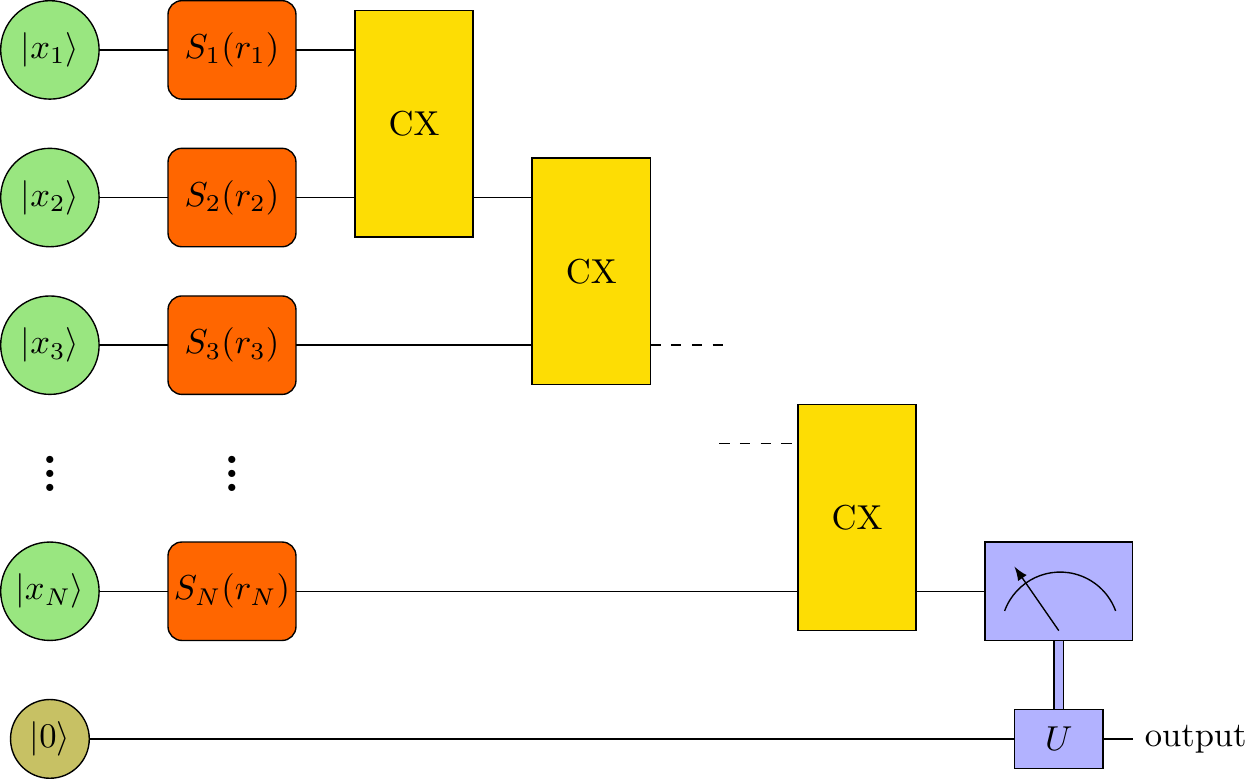}
	\caption{Scheme for a continuous valued quantum perceptron}
	\label{fig:QN}
\end{figure}
We will now use the classical statistical approach {outlined above} to compute the pattern capacity of the particular quantum perceptron introduced in~\cite{BenattiManciniMangini} and schematically represented by the quantum circuit in Fig.~\ref{fig:QN}.
In this model, {the components $x_j$ of an input pattern $\xv$ are encoded by states of the form 
	\begin{equation}
		\vert \psi_j\rangle = \frac{1}{(2\pi \sigma^2_j)^{1/4}} \int_{-\infty}^{+\infty} d q_j \ {   \exp\left(-\frac{(q_j - x_j)^2}{4 \sigma^2_j}\right)}\vert q_j\rangle\ ,
		\label{eq:WavePacket}
	\end{equation}
	which are Gaussian weighted normalized superpositions of  pseudo-eigenstates $\vert q_j\rangle$ of position-like operators $q$, centered around the pattern components $x_j$ with width $\sigma_j$. 
	A pattern $\xv$ is then encoded into the tensor product state
	\begin{equation}
		\vert \Psi\rangle=\bigotimes_{j=1}^N\vert \psi_j\rangle=\prod_{j=1}^N \frac{1}{(2\pi \sigma_j^2)^{1/4}}\int_{-\infty}^{+\infty} dq_j\,{  \exp\left(-\frac{(q_j - x_j)^2}{4 \sigma^2_j}\right)}
		\bigotimes_{j=1}^N\vert q_j\rangle\ .
		\label{eq:Nwavepacket}
	\end{equation}
	Such state is then provided as input to the quantum circuit in Fig.~\ref{fig:QN}, which firstly operates with a series of independent squeezing
	operators
	\begin{equation}
		\label{Sq1}
		S_j(r_j)=\rme^{\rmi\,r_j\,(q_jp_j+p_jq_j)}\ ,\quad {\rm e}^{-2r_j}=w_j
	\end{equation} 
	where $p_j$ is a momentum-like operator conjugated to $q_j$, $[q_j,p_j]=\rmi$. 
	In order to implement negative weights, a phase shift gate $\rme^{\rmi\frac{\pi}{2}(q^2+p^2)}$ is operated after the squeezing, thus sending position 
	eigenstates $\vert q_j\rangle$ into $\vert w_j\,q_j\rangle$.
	
	Then, the circuit acts with an entangling Controlled Addition gate $\CX$ on pairs of consecutive states:
	\begin{equation}
		\label{eq:CX}
		\CX:= \exp\left(-\rmi\, q_{j} \otimes p_{j+1}\right)\ ,\qquad \CX\,\vert q_j, q_{j+1}\rangle=\vert q_j,q_j+q_{j+1}\rangle\ .
	\end{equation}
	The combined action on the attenuated multi-mode  position eigen-state of the $n-1$ $\CX$ gates of the circuit in Fig.~\ref{fig:QN}  is then given by
	\begin{equation}
		\begin{split}
			\vert w_1 q_1, w_2 q_2, \hdots,w_n q_n\rangle & \rightarrow \vert w_1 q_1, w_1 q_1+w_2 q_2, \hdots,w_n q_n\rangle  \rightarrow \hdots \\ 
			& \hdots \rightarrow \vert w_1 q_1,w_1 q_1+w_2 q_2, \hdots,\sum_{j=1}^{N} w_j q_j\rangle\ .
		\end{split}
	\end{equation}
Then, the amplitude associated to the last mode position eigenstate is given by the Gaussian distribution centered around $\wv\cdot\xv^\mu$:
		\begin{equation}
			\psi_{\wv,\xv^{\mu}}(s)=\frac{1}{(2\pi\sum_j w_j^2 \sigma_j^2)^{1/4}}\exp\left(-\frac{(s-\wv\cdot \xv^\mu)^2}{4\sum_{j}w_j^2\sigma_j^2}\right).
		\end{equation}
The variances $\sigma_j^2$ can be suitably chosen. Here, we will consider the case $\sigma_j^2=\sigma^2$ for all $j$.
Homodyne detection operated on the last mode yields the outcome probability 
		\begin{equation}\label{eq:Prob}
			P_{\wv,\xv^{\mu},\sigma}(s)=|\psi_{\wv,\xv^{\mu}}(s)|^2=\frac{1}{\sqrt{2\pi}\|\wv\|\sigma}\exp\left(-\frac{(s-\wv\cdot \xv^\mu)^2}{2\|\wv\|^2\sigma^2}\right).
		\end{equation}

Consequently a pattern $\xv^\mu$ is classified as $\sigma=+1$ if the result is above some threshold $\kappa\|\wv\|$, and $\sigma=-1$ if it is below $-\kappa\|\wv\|$, while the pattern is not classified when the measurement result is between $(-\kappa\|\wv\|,\kappa\|\wv\|)$. Then, {  whether a pattern is correctly classified or not becomes a binary stochastic  variable for which} the probability of correct classification of the pattern $\mu$ is 
	\begin{equation}\label{eq:rmu}
		R^\mu={  \int_{-\infty}^{+\infty}} \de s~ P_{\wv,\xv^{\mu},\sigma}(s)\theta\left(\xi^\mu\frac{s}{\|\wv\|}-\kappa\right)
	\end{equation}
Depending on the actual outcome of the measurement process, the ancilla mode appended to the initialized first $N$ ones will then be changed accordingly. For this classification problem the ancilla can be taken as a qutrit, which is initialized in some reference state, while two orthogonal states are used to store respectively the classifications $\sigma=+1$ and $\sigma=-1$.
	
\begin{remark}
\label{rem-class}
{  Notice that the classical perceptron action is exactly encoded by the position-like pseudo-autokets $\vert q_1,q_2,\ldots,q_N\rangle$. However, such an ideal situation 
can only be approximated in practice. One useful way to achieve it is via Gaussian smoothening, the narrowing of the weights thus corresponding to the classical deterministic limit of the proposed continuous quantum perceptron.}
\end{remark}
	
\subsection{Results}
	Notice that a pattern which would be correctly classified by a classical perceptron (i.e. a pattern for which $\Delta^\mu>0$) has a non-vanishing probability of being misclassified by such a model of quantum perceptron, namely $R^\mu<1$ even if $\Delta^\mu>\kappa$. Viceversa, even if the classical perceptron misclassifies $\xi^\mu$ yielding $\Delta^\mu<\kappa$, the quantum one would classify it correctly with some probability. This kind of error comes from the random character of the outcomes $s$ of the homodyne measurement which yields the value $s$ either in $(-\infty,\kappa\,\|\wv\|]\cup[\kappa\,\|\wv\|,\infty)$ or in $(-\kappa\,\|\wv\|,\kappa\,\|\wv\|)$. 
	Because of the unavoidable presence of the stochastic error due to the quantum setting and {  the} corresponding rendering of the perceptron non-linearity by a homodyne quantum measurement, the computation by Gardner requires the introduction of an additional parameter $\epsilon$ establishing a threshold for the acceptable probability of error in the classification of each pattern. Then, the relative volume we are interested in is given by
	\begin{equation}\label{eq:relVq}
		V:=V(\{\xv^{\mu},\xi^\mu\}_{\mu=1}^p)=\frac{1}{C_N}{  \int_{\mathbb{R}^N}} \de \wv~ \delta(\|\wv\|^2-N) \prod_{\mu=1}^p\theta \left(R^\mu-1+\epsilon\right).
	\end{equation}
	\begin{remark}
		In equation \eqref{eq:relVq} we are computing the volume in the space of weights which guarantees that the probability of making an error \textit{on each} pattern is smaller than $\epsilon$, {  that will then be chosen smaller than $1/2$}. This does not imply that {  $\epsilon p\leq p/2$} errors will occur on average for the $p$ patterns, but rather that {  $\epsilon p\leq p/2$} is an upper bound to the number of the patterns which will be misclassified by the perceptron.
	\end{remark}
	As before, the statistically relevant quantity is $\left\langle \ln V\right \rangle$, which can be computed using the replica method {  together with} the replica symmetric ansatz, as detailed in the next section. We obtain that the pattern capacity of the quantum model presented is given by the critical value
	\begin{equation}\label{eq:critCapQ}
		\alpha_c^q(\kappa,\epsilon,\sigma)=\alpha_c(\widetilde{\kappa}),
	\end{equation}
	where
	\begin{equation}\label{eq:kappatilde}
		\widetilde{\kappa}=\kappa+ \sigma \Phi^{-1}(1-\epsilon),
	\end{equation}
	In particular, for $\alpha<\alpha_c(\widetilde{\kappa})$ the following limit holds:
\begin{equation}\label{eq:aveLogV1Q}
	\lim_{\stackrel{N,p\rightarrow \infty}{{  p/N=\alpha}}}\frac{\langle \ln V \rangle}{N} = \mathcal{F}(\alpha,\widetilde{\kappa})
\end{equation}
where $\mathcal{F}$ is the same as in equation \eqref{eq:aveLogV1}, while for $\alpha> \alpha_c(\kappa)$
\begin{equation}\label{eq:aveLogV2Q}
	\lim_{\stackrel{N,p\rightarrow \infty}{{  p/N=\alpha}}}\frac{\langle \ln V \rangle}{N} = - \infty.
\end{equation}

Since $\alpha_c(\kappa)$ is decreasing in $\kappa$, equation~\eqref{eq:kappatilde} implies that (recall that $0<\epsilon<1/2$) the quantum pattern capacity is smaller than its classical counterpart (see also \figurename{~\ref{fig:critcapacity}).
	\begin{figure}[h]
		\includegraphics[width=.49\textwidth]{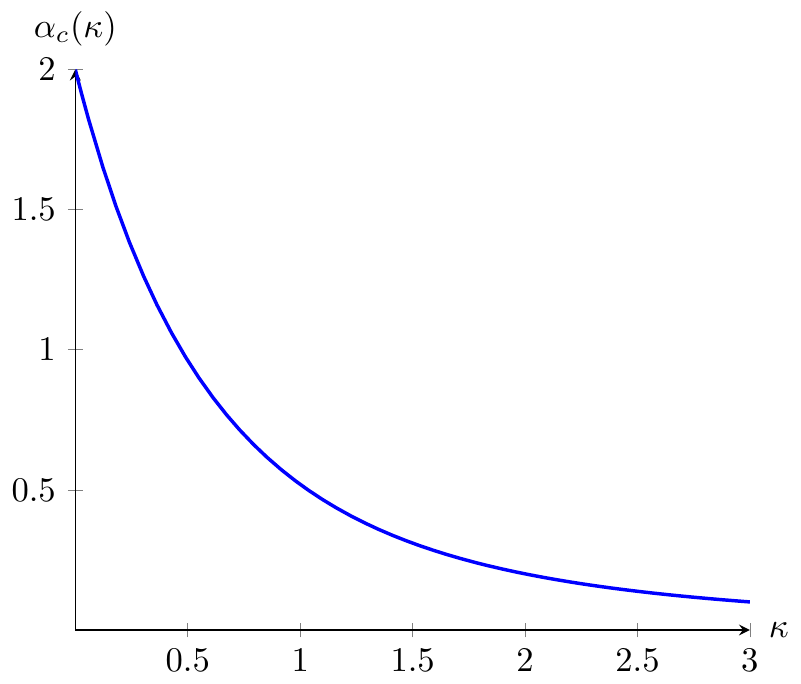}~
		\includegraphics[width=.49\textwidth,trim=0 -18pt 0 0]{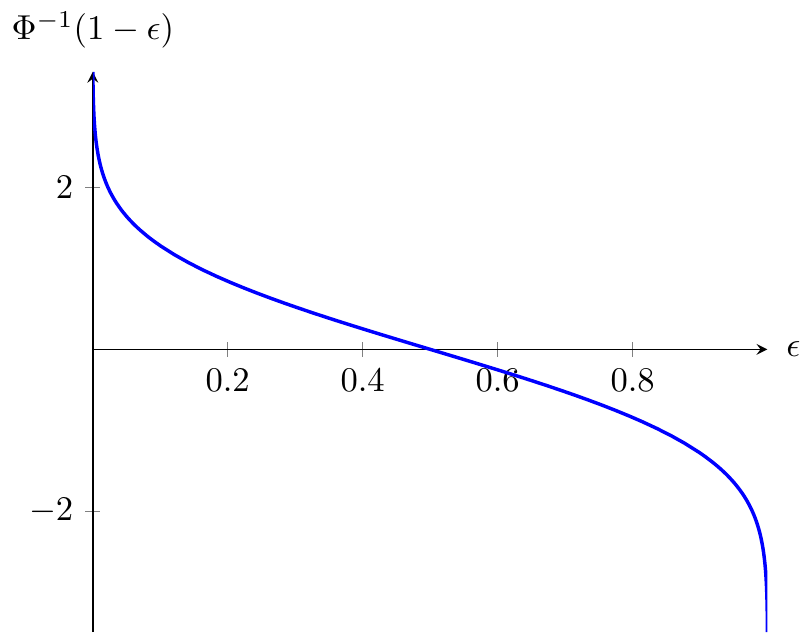}
		\caption{$\alpha_c(\kappa)$ is monotonically decreasing in $\kappa$, and $\Phi^{-1}(1-\epsilon)$ is positive for $0<\epsilon<1/2$. Then, for $0<\epsilon<1/2$, $\alpha_c^q(\kappa,\epsilon,\sigma)<\alpha_c(\kappa)$.}
		\label{fig:critcapacity}
	\end{figure}

\begin{remark}
Note that, {   according to Remark~\ref{rem-class}, the functioning of the classical perceptron is recovered} from the quantum model proposed by letting 
$\sigma\rightarrow 0$, {  so that \eqref{eq:Prob} gives} $\lim_{\sigma\rightarrow0}P_{\wv,\xv^{\mu},\sigma}=\delta(s-\wv\cdot \xv)$, {  and} $\widetilde{\kappa}\rightarrow \kappa$, 
as it should be.
\end{remark}

\section{Methods}\label{sec:Methods}
In the following we present the derivation of equations \eqref{eq:critCapQ}.
{ Using the replica trick}, one has to compute
\begin{equation}
	\langle \ln V \rangle =\lim_{n\rightarrow 0} \frac{\langle V^n\rangle -1}{n}.
\end{equation}
by evaluating $\langle V^n\rangle$ for $n$ integer, and then taking the limit $n\rightarrow 0$. The quantity we need is then:
\begin{equation}\label{eq:aveVn}
	\langle V^n \rangle =\frac{1}{C_N^n}\left\langle\prod_{\gamma=1}^n\int_{{ \mathbb{R}^N}} \de \wv^\gamma~ \delta(\|\wv^\gamma\|^2-N)  \prod_{\mu=1}^p\theta \left(R^\mu_\gamma(\kappa)-1+\epsilon\right)\right\rangle.
\end{equation}
Using the integral representation 
\begin{equation}
	\theta(z-\kappa)=\int_{\kappa}^{\infty} \de \lambda~\delta(\lambda-z)=\int_{\kappa}^{\infty}\de \lambda \int \frac{\de y}{2\pi} \rme^{\rmi y(\lambda-z)},
\end{equation}
one writes
\begin{equation}
	\theta \left(R^\mu_\gamma(\kappa) -1+\epsilon\right)=\int_{1-\epsilon}^\infty \frac{\de z_\gamma^\mu}{2\pi} \int_{-\infty}^{\infty} \de y_\gamma^\mu ~\rme^{\rmi y_\gamma^\mu z_\gamma^\mu}\rme^{-\rmi y_\gamma^\mu R^\mu_\gamma(\kappa)}.
\end{equation}
Now resorting to the fact that each pattern is independent from the other ones, we can factor the average into:
\begin{align}\notag
	&\left\langle \prod_{\mu=1}^p\prod_{\gamma=1}^n \theta \left(R^\mu_\gamma(\kappa) -1+\epsilon\right)\right\rangle \\
	\notag
	&\qquad =\prod_{\mu=1}^p\left\langle\prod_{\gamma=1}^n \theta \left(R^\mu_\gamma(\kappa) -1+\epsilon\right)\right\rangle\\
	&\qquad =\prod_{\mu=1}^p\int_{[1-\epsilon,\infty)^n} \left(\prod_{\gamma=1}^n\frac{\de z_\gamma^\mu}{2\pi}\right) \int_{\mathbb{R}^n} \left(\prod_{\gamma=1}^n\de y_\gamma^\mu\right) ~\rme^{\rmi \sum_{\gamma=1}^n y_\gamma^\mu z_\gamma^\mu}\left\langle\prod_{\gamma=1}^n\rme^{-\rmi y_\gamma^\mu R^\mu_\gamma(\kappa)}\right\rangle
	\label{eq:avethetaInt}
\end{align}
Then, we have to go through the computation of:
\begin{equation}
	\left\langle \prod_{\gamma=1}^n \rme^{-\rmi y_\gamma^\mu R^\mu_\gamma(\kappa)}\right\rangle_{\Pi,\xiv},
\end{equation}
where we wrote explicitly the variables over which we are averaging. 
The details of the calculation are given in Appendix~\ref{app:aveComp}, the final result being:
\begin{align}\notag
	&\left\langle \prod_{\mu=1}^p\prod_{\gamma=1}^n \theta \left(R^\mu_\gamma(\kappa) -1+\epsilon\right)\right\rangle \\
	&\qquad = \left[\int_{[1-\epsilon,\infty)^n} \left(\prod_{\gamma=1}^n\frac{\de z_\gamma}{2\pi}\right)\int_{\mathbb{R}^n\times\mathbb{R}^n\times \mathbb{R}^n} \left(\prod_{\gamma=1}^n \frac{\de \lambda_{\gamma} \de y_\gamma \de \omega_{\gamma}}{2\pi}\right)\rme^{K(\{\lambda_\gamma\},\{y_\gamma\},\{\omega_\gamma\},\{q_{\gamma\delta}\})}\right]^p,
	\label{eq:avethetaFinal}
\end{align}
where 
\begin{equation}\label{eq:K}
	K(\{\lambda_\gamma\},\{y_\gamma\},\{\omega_\gamma\},\{q_{\gamma\delta}\})\equiv\rmi \sum_{\gamma=1}^n y_\gamma \left[z_\gamma- \Phi(\lambda_\gamma)\right] -\rmi \sum_{\gamma=1}^n \left(\frac{\kappa}{\sigma}+\lambda_\gamma\right)\omega_{\gamma}^{\mu}-\frac{1}{2\sigma^2 }\sum_{\gamma,\delta=1}^n q_{\gamma\delta} \omega_\gamma\omega_\delta
\end{equation}
and 
\begin{equation}\label{eq:qdef}
	q_{\gamma \delta}=\frac{1}{N}\sum_{j=1}^{N} w_j^\gamma w_j^\delta, \quad \text{for }\gamma,\delta=1,\dots,n,
\end{equation}
(satisfying $q_{\gamma\gamma}=1$ due to the constraint $\|\wv^\gamma\|=\sqrt{N}$) are order parameters introduced by means of
{delta functions,}
\begin{equation}\label{eq:deltaq0}
	\delta\bigg(q_{\gamma\delta}-\frac{1}{N}{\sum_{j=1}^N}\,w_j^\gamma w_j^\delta\bigg)=N{ \int_{-\infty}^{+\infty}} \frac{\de F_{\gamma \delta}}{2\pi} \exp\bigg(-\rmi Nq_{\gamma\delta}F_{\gamma \delta}+\rmi F_{\gamma\delta} \sum_{j=1}^N w_j^\gamma w_j^\delta\bigg),
\end{equation}
which involves the introduction of further parameters $F_{\gamma\delta}$, for $\gamma\neq \delta$. Similarly, from the integral representation 
\begin{equation}\label{eq:deltaNorm}
	\delta\bigg({\sum_{j=1}^N} (w_j^\gamma)^2-N\bigg)= {\int_{-\infty}^{+\infty}} \frac{\de E_\gamma}{4\pi}\, {\exp\bigg[\rmi \frac{E_\gamma}{2}\bigg(N-\sum_{j=1}^N\,(w_j^\gamma)^2\bigg)\bigg]},
\end{equation}
we get further order parameters $E_\gamma$, for $\gamma=1,\dots,n$. Then, $\langle V^n\rangle$ can be globally rewritten as:
\begin{equation}\label{eq:aveVnBis}
	\langle V^n \rangle =\frac{1}{C_N^n}\int_{ \mathbb{R}^n\times\mathbb{R}^{n(n-1)/2}\times\mathbb{R}^{n(n-1)/2}}\Big(\prod_{\gamma{ =1}}^{ n} \de E_\gamma\Big)\Big(\prod_{\substack{\gamma,\delta=1\\ \gamma < \delta}}^{ n}\,\de q_{\gamma \delta}\,\de F_{\gamma \delta}\Big)~\rme^{NG(\{q_{\gamma\delta}\},\{F_{\gamma\delta}\},\{E_{\gamma}\})}
\end{equation}
where 
\begin{align}\label{eq:G}
	G(\{q_{\gamma\delta}\},\{F_{\gamma\delta}\},\{E_{\gamma}\})=\alpha G_1(\{q_{\gamma\delta}\})+G_2(\{F_{\gamma\delta}\},\{E_{\gamma}\})+G_3(\{q_{\gamma\delta}\},\{F_{\gamma\delta}\},\{E_{\gamma}\}),
\end{align}
with
\begin{align}\label{eq:G10}
	&G_1(\{q_{\gamma\delta}\})=\ln \left[\int_{[1-\epsilon,\infty)^n} \left(\prod_{\gamma=1}^n\frac{\de z_\gamma}{2\pi}\right)\int_{\mathbb{R}^n\times\mathbb{R}^n\times \mathbb{R}^n} \left(\prod_{\gamma=1}^n \frac{\de \lambda_{\gamma} \de y_\gamma \de \omega_{\gamma}}{2\pi}\right)\rme^{K(\{\lambda_\gamma\},\{y_\gamma\},\{\omega_\gamma\},\{q_{\gamma\delta}\})}\right]\\
	\label{eq:G20}
	&G_2(\{F_{\gamma\delta}\},\{E_{\gamma}\})=\ln \Bigg[\int_{\mathbb{R}^n} \Bigg({ \prod_{\gamma=1}^n} \de w^\gamma \Bigg) 
	\exp\Bigg(-\frac{\rmi}{2}{\sum_{\gamma=1}^n} E_\gamma (w^\gamma)^2+{\rmi \sum_{\substack{\gamma,\delta=1 \\ \gamma<\delta}}^n} F_{\gamma \delta}w^\gamma w^\delta\Bigg)\Bigg]\\
	\label{eq:G30}
	&G_3(\{q_{\gamma\delta}\},\{F_{\gamma\delta}\},\{E_{\gamma}\})=-{ \rmi \sum_{\substack{\gamma,\delta=1\\ \gamma< \delta}}^n} 
	F_{\gamma \delta}q_{\gamma\delta}+\frac{\rmi}{2}{\sum_{\gamma=1}^n}E_\gamma.
\end{align}
When $N$ is large, the behaviour of $\langle V^n\rangle$ 
can be obtained using the saddle-point approximation:
\begin{equation}\label{eq:aveVnSP}
	\langle V^n \rangle { \simeq} \frac{1}{C_N^n}\rme^{N G(z_S)}\sqrt{\frac{2\pi}{N|\det G''(z_S)|}},
\end{equation}
where $z_S=(\{q_{\gamma\delta}^S\},\{F_{\gamma\delta}^S\},\{E_\gamma^S\})$ is the stationary point of $G$, and $G''(z_S)$ denotes the Hessian matrix of $G$ at the stationary point.

Using \eqref{eq:aveVnSP}, and the fact that $\langle V^n\rangle \rightarrow 1$ as $n\rightarrow 0$, we can write:
\begin{align}\notag
	\frac{\langle \ln V \rangle }{N}&=\lim_{n\rightarrow 0}\frac{\langle V^n\rangle -1}{nN}\\\notag
	&=\lim_{n\rightarrow 0}\frac{\ln \langle V^n\rangle}{nN}\\\notag
	& = \lim_{n\rightarrow 0}\frac{1}{n}\left[G(z_S)-\frac{\ln C_N^n}{N} +\frac{1}{2N}\ln\left(\frac{2\pi}{N|\det G''(z_S)|}\right) \right]\\
	&=\lim_{n\rightarrow 0}\frac{G(z_S)}{n}-\frac{\ln (2\pi \rme)}{2} +O\left(\frac{\ln N}{N}\right).
	\label{eq:avelnV/N}
\end{align}
\subsection{Replica-symmetric ansatz}
At this point the replica-symmetric ansatz is made, which consists in the assumption that the stationary point of $G$ can be found among the points $z=(\{q_{\gamma\delta}\},\{F_{\gamma\delta}\},\{E_\gamma\})$ such that:
\begin{equation}\label{eq:replicaSymmetryAnsatz}
	q_{\gamma \delta}=q \qquad F_{\gamma \delta}=F \qquad E_\gamma=E,
\end{equation}
for all $\gamma,\delta=1,\dots,n$, with $\gamma\neq \delta$. In the following, we denote with 
$$
	G^{\RS}(q,F,E)=\alpha G_1^{\RS}(q)+G_2^{\RS}(F,E)+G_3^{\RS}(q,F,E)
$$
the restriction of $G$ onto the replica-symmetric subspace. 
The details of the derivation of $G_1^{\RS}(q)$ are given in Appendix~\ref{app:G1RS}, which yields:
\begin{equation}
	\label{eq:G1RS}
	G_1^{\RS}=\ln\left[\int_{-\infty}^{\infty}\de t\frac{\rme^{-t^2/2}}{\sqrt{2\pi}}\left(1-\Phi\left(\frac{\widetilde{\kappa}+t\sqrt{q}}{\sqrt{1-q}} \right)\right)^n \right].
\end{equation}
In the limit $n\rightarrow 0$ we obtain the asymptotic expansion
\begin{equation}
	\label{eq:G1RSasy}
	G_1^{\RS}=n\int_{-\infty}^{\infty}\de t\frac{\rme^{-t^2/2}}{\sqrt{2\pi}}\ln \left(1-\Phi\left(\frac{\widetilde{\kappa}+t\sqrt{q}}{\sqrt{1-q}} \right)\right)+O(n^2),
\end{equation}
which follows from the fact that for $n\rightarrow 0$:
\begin{align}\notag
	\ln\left[\int_{-\infty}^{\infty}\de t\frac{\rme^{-t^2/2}}{\sqrt{2\pi}}f(t)^n \right]&=
	\ln \left[1+\int_{-\infty}^{\infty}\de t\frac{\rme^{-t^2/2}}{\sqrt{2\pi}}(f(t)^n-1) \right]\\
	\notag
	&=\int_{-\infty}^{\infty}\de t\frac{\rme^{-t^2/2}}{\sqrt{2\pi}}(f(t)^n-1)+O(n^2)\\
	&=n\int_{-\infty}^{\infty}\de t\frac{\rme^{-t^2/2}}{\sqrt{2\pi}}\ln(f(t))+O(n^2).
\end{align}
The derivation of $G_2^\RS$ follows simply by a gaussian integration:
\begin{equation}
	G_2^{\RS}(F,E)=\frac{1}{2}\ln\left(\frac{(2\pi \rmi)^n}{\det \Lambda} \right),
\end{equation}
where $\Lambda_{\gamma\delta}=-(E+F)\delta_{\gamma\delta}+F$.
The determinant is easily computed by noticing that the matrix $\Lambda$ has one non-degenerate eigenvalue $\lambda_1=-E+nF$ and a $(n-1)$-degenerate eigenvalue $\lambda_2=-(E+F)$, so that we can write
\begin{align}\notag
	G_2^\RS &= \frac{1}{2}\ln\left[ \frac{(2\pi \rmi)^n}{(-1)^n (E+F)^{n-1}(E-(n-1)F)} \right]\\
	&=\frac{n}{2}\left( \ln(2\pi)-\ln( \rmi E+ \rmi F)+\frac{F}{E+F}\right)+O(n^2).
	\label{eq:G2RS}
\end{align}
Finally, the last term is simply given by;
\begin{equation}\label{eq:G3RS}
	G_3^{\RS}(q,F,E)=\rmi \frac{n}{2}(E+qF)+O(n^2).
\end{equation}
Therefore, in the limit $n\rightarrow 0$ we have:
\begin{align}\notag
	&\frac{1}{n}G^{\RS}(q,F,E)=\\
	&\quad=\alpha \int_{-\infty}^{+\infty} \frac{\de t}{\sqrt{2\pi}} \rme^{-t^2/2}\ln\left[1-\Phi\left(\frac{t\sqrt{q}+\widetilde{\kappa}}{\sqrt{1-q}}\right) \right]+\frac{1}{2}\left( \ln(2\pi)-\ln(\rmi E+\rmi F)+\frac{F}{E+F}+\rmi E+\rmi qF\right).
	\label{eq:GRS}
\end{align}
The saddle point equations $\frac{\partial G^{\RS}}{\partial E}=0$ and $\frac{\partial G^{\RS}}{\partial F}=0$ can readily be solved, giving
\begin{equation}\label{eq:FES}
	F_S(q)=-\frac{\rmi q}{(1-q)^2}, \qquad E_S(q)=-\rmi  \frac{1-2q}{(1-q)^2},
\end{equation}
which, 
{ substituted into $G^{\RS}$, yield:}
\begin{align}\notag
	&\frac{1}{n}G^{\RS}(q,F_S(q),E_S(q))=\\
	&\quad=\alpha \int_{-\infty}^{+\infty} \frac{\de t}{\sqrt{2\pi}} \rme^{-t^2/2}\ln\left[1-\Phi\left(\frac{t\sqrt{q}+\widetilde{\kappa}}{\sqrt{1-q}}\right) \right]+\frac{\ln(2\pi)}{2}+\frac{\ln(1-q)}{2}+\frac{q}{2(1-q)}+\frac{1}{2}.
	\label{eq:G(q)0}
\end{align}
Setting $\partial G^{\RS}/\partial q=0$ in \eqref{eq:G(q)0} we get the saddle point equation
\begin{equation}\label{eq:SaddlePointQ}
	\alpha \int \frac{\de t}{\sqrt{2\pi}}\rme^{-t^2/2}\mathcal{A}\left(\frac{ \widetilde{\kappa}+t \sqrt{q}}{\sqrt{1-q}}\right)\frac{t+ \widetilde{\kappa}\sqrt{q}}{2\sqrt{q}(1-q)^{3/2}}=\frac{q}{2(1-q)^2},
\end{equation}
where 
\begin{equation}
	\mathcal{A}(u):=\frac{1}{\sqrt{2\pi}}\frac{\rme^{-u^2/2}}{1-\Phi(u)}.
\end{equation}
The most important order parameter is $q$, whose value at the replica symmetric-saddle point represents the most probable average overlap \eqref{eq:qdef} between a pair of solutions to \eqref{eq:CondStabBis}. For $\alpha\rightarrow0$, equation \eqref{eq:SaddlePointQ} gives $q\rightarrow0$: in this case almost all $\wv$'s solve \eqref{eq:CondStabBis}, and the typical overlap between random pairs of $\wv$ in the space of interactions is vanishing. As $\alpha$ grows, it becomes harder and harder to find solutions, hence the typical overlap between them increases. The optimal perceptron corresponds to the limit $q\rightarrow 1$, when there is only a single solution solving the problem with the given stability $\kappa$, and the corresponding value of $\alpha$ is the critical storage capacity $\alpha_c^q(\kappa,\epsilon,\sigma)$.
The critical value of $\alpha_c^q(\kappa,\epsilon,\sigma)$ is obtained thus by Equation~\eqref{eq:SaddlePointQ} taking the limit $q\rightarrow 1$:
\begin{equation}\label{eq:alpha_c0}
	\frac{1}{\alpha_c^q(\kappa,\epsilon,\sigma)}=\lim_{q\rightarrow 1}\int \frac{\de t}{\sqrt{2\pi}} \rme^{-t^2/2}\mathcal{A}\left(\frac{ \widetilde{\kappa}+t \sqrt{q}}{\sqrt{1-q}}\right)\frac{\sqrt{1-q}}{q^{3/2}}(t+ \widetilde{\kappa} \sqrt{q}),.
\end{equation}
Using the asymptotic expansion
\begin{equation}
	\mathcal{A}(u)= u+O\left(\frac{1}{u}\right) \qquad \text{for }u\rightarrow \infty
\end{equation}
and $\mathcal{A}(u)\rightarrow 0$ as $u\rightarrow -\infty$, one obtains that
\begin{equation}
	\lim_{q\rightarrow 1}\mathcal{A}\left(\frac{ \widetilde{\kappa}+t \sqrt{q}}{\sqrt{1-q}}\right)\frac{\sqrt{1-q}}{q^{3/2}}(t+ \widetilde{\kappa} \sqrt{q})= (t+ \widetilde{\kappa})^2 \theta(t+ \widetilde{\kappa}).
\end{equation}
Then, by the dominated convergence theorem, Equation~\eqref{eq:alpha_c0} gives the final result:
\begin{equation}
	\alpha_c^q(\kappa,\epsilon,\sigma)=\left[\int_{- \widetilde{\kappa}}^{\infty}\frac{\de t}{\sqrt{2\pi}}\rme^{-t^2/2}(t+ \widetilde{\kappa})^2 \right]^{-1}=\alpha_c( \widetilde{\kappa}).
\end{equation}

\section{Discussion}\label{sec:Conclusion}

In order to  investigate possible quantum advantages of
quantum machine learning with respect to its classical counterpart 
the design of the elementary blocks of quantum neural networks, namely of quantum perceptrons, is of primary importance. 
In  this work, we have considered a particular model of  continuous variable quantum perceptron and showed that in this case there is no quantum advantage as far as its pattern capacity is concerned, when compared to the classical one. 
The main reason for this conclusion is that the quantum model closely resembles a stochastic classical perceptron due to the uncertainty in the outcomes of the 
(homodyne) measurements necessary to extract the relevant information, namely the stabilities $\Delta^\mu$, that is the linear combinations of pattern inputs and weights.
The statistical, replica-trick techniques adopted in our investigation are however quite flexible, so that they could be possibly employed to explore {different and less classically inspired} models of quantum perceptrons. 

Indeed, upon completion of this work, we discovered that one such attempt has been recently made in \cite{Lewenstein21}. The model considered there is {  a discrete variable model} based on the encoding of {  the stabilities as transition amplitudes between suitably constructed quantum state vectors}. The activation function is then evaluated on the measured value of the transition probability $\left|\Delta^\mu\right|^2$. Instead, 
in our case we infer the value of  $\mathrm{sgn}(\Delta^\mu)$ by measuring a gaussianly distributed parameter (see \eqref{eq:Prob} and the corresponding discussion), which mimics the functioning of a stochastic classical perceptron. Using the {  same} statistical techniques employed here, the authors find that the pattern capacity of that model is twice the classical capacity. The improvement seemingly arises from the fact that the transition probability $\left|\Delta^\mu\right|^2$ does not distinguish the sign of the stabilities which are instead distinguished by $\mathrm{sgn}(\Delta^\mu)$. 
However, differently from the case studied in this work, the use of the measured probabilities makes somewhat difficult 
to provide a classical limit for this discrete-variable quantum perceptron with which to compare the advantages it brings about.

In this respect, in order to better understand whether advantages can be expected and how they can be achieved, an avenue certainly to be explored is the more information-rooted approach~\cite{Brunel} that considers the information stored in a perceptron rather than the number of patterns.

\section*{Acknowledgements}
F.B. and S.M. warmly thank Samad Khabbazi-Oskouei for useful discussions in the early stage of this work. G.G. is partially
supported by the Italian National Group of Mathematical Physics (GNFM-INdAM).

\appendix

\section{Derivation of \eqref{eq:avethetaFinal}}
\label{app:aveComp}
 Since $x_j^\mu$ and $\xi^\mu$ are independent random variables, the average over the $\xi^\mu$'s can be carried out immediately, yielding:
\begin{align}\notag
	\left\langle \prod_{\gamma=1}^n \rme^{-\rmi y_\gamma^\mu R^\mu_\gamma(\kappa)}\right\rangle_{\Pi,\xiv}& =\frac{1}{2}\Bigg[\left\langle\prod_{\gamma=1}^n \exp\left(-\rmi y_\gamma^\mu \int_{\kappa\|\wv^\gamma\|}^\infty \de s~ P_{\wv^\gamma,\xv^{\mu}}(s)\right)\right\rangle_{\Pi}\\
	&\qquad \qquad   +\left\langle\prod_{\gamma=1}^n \exp\left(-\rmi y_\gamma^\mu \int^{-\kappa\|\wv^\gamma\|}_{-\infty} \de s~ P_{\wv^\gamma,\xv^{\mu}}(s)\right)\right\rangle_{\Pi}\Bigg],
	\label{eq:aveQuantumTheta0}
\end{align}
Now noting that
\begin{align}
	\int_{\kappa\|\wv\|}^\infty \de s~ P_{\wv,\xv}(s)&=1-\Phi\left( \frac{\kappa}{\sigma}-\frac{\wv\cdot \xv}{\|\wv\|\sigma}\right), \\
	\int_{-\infty}^{-\kappa\|\wv\|}\de s~ P_{\wv,\xv}(s)&=\Phi\left( -\frac{\kappa}{\sigma}-\frac{\wv\cdot \xv}{\|\wv\|\sigma}\right),
\end{align}
using the fact that $1-\Phi\left(x\right)=\Phi\left( -x\right)$, and recalling that we are integrating over the sphere with $\|\wv\|=\sqrt{N}$, we have
\begin{align}
	\notag
	&\left\langle\prod_{\gamma=1}^n \rme^{-\rmi y_\gamma^\mu R^\mu_\gamma(\kappa)}\right\rangle_{\Pi,\xiv}\\
	\label{eq:aveQuant0}
	&\quad =\frac{1}{2}\left[\left\langle \prod_{\gamma=1}^n\exp\left(-\rmi y_\gamma^\mu \Phi\left(\frac{\wv^\gamma\cdot\xv^\mu}{\sigma \sqrt{N}}-\frac{\kappa}{\sigma}\right)\right)\right\rangle_{\Pi} +\left\langle \prod_{\gamma=1}^n\exp\left(-\rmi y_\gamma^\mu \Phi\left(-\frac{\kappa}{\sigma}-\frac{\wv^\gamma\cdot\xv^\mu}{\sigma \sqrt{N}}\right)\right)\right\rangle_{\Pi}\right]
\end{align}
In order to compute the averages of the two terms inside the square brackets, we write the functions $f_y(\lambda)=\rme^{-\rmi y \Phi(\lambda)}$ in terms of their fourier transforms:
\begin{equation}\label{eq:fourier}
	f_y(\lambda)=\frac{1}{\sqrt{2\pi}}\int \de \omega ~\rme^{\rmi \omega  \lambda}\hat{f}_y(\omega), \qquad \hat{f}_y(\omega)=\frac{1}{\sqrt{2\pi}}\int \de t~ \rme^{-\rmi \omega t}f_y(\lambda).
\end{equation}
We can then write:
\begin{align}\notag
	&\left\langle\prod_{\gamma=1}^n \exp\left(-\rmi y_\gamma^\mu \Phi\left(\frac{\wv^\gamma\cdot\xv^\mu}{\sigma \sqrt{N}} -\frac{\kappa}{\sigma}\right)\right)\right\rangle_{\Pi}\\
	\notag
	&\qquad=\frac{1}{(2\pi)^{n/2}}\int_{\mathbb{R}^n} \Bigg( \prod_{\gamma=1}^n\de \omega_{\gamma}^\mu \Bigg)\left\langle \exp\Bigg(\rmi\sum_{\gamma=1}^n \omega_{\gamma}^\mu\frac{\wv^\gamma\cdot\xv^\mu}{\sigma \sqrt{N}}\Bigg) \right\rangle_{\Pi}\rme^{-\rmi \frac{\kappa}{\sigma}\sum_{\gamma=1}^n \omega_{\gamma}^{\mu}}\prod_{\gamma=1}^n \hat{f}_{y_\gamma^\mu}(\omega_{\gamma}^\mu)\\
	\notag
	&\qquad=\frac{1}{(2\pi)^{n/2}}\int_{\mathbb{R}^n} \Bigg( \prod_{\gamma=1}^n\de \omega_{\gamma}^\mu \Bigg) \exp\left[\sum_{j=1}^N \ln\cos\left(\frac{1}{\sigma \sqrt{N}}\sum_{\gamma=1}^n w_j^\gamma \omega_\gamma^\mu\right)\right]\rme^{-\rmi \frac{\kappa}{\sigma}\sum_{\gamma=1}^n\omega_{\gamma}^{\mu}}\prod_{\gamma=1}^n\hat{f}_{y_\gamma^\mu}(\omega_{\gamma}^\mu)\\
	&\qquad\simeq\frac{1}{(2\pi)^{n/2}}\int_{\mathbb{R}^n} \Bigg( \prod_{\gamma=1}^n\de \omega_{\gamma}^\mu \Bigg)  \exp\Bigg(-\frac{1}{2\sigma^2 }\sum_{\gamma,\delta=1}^n q_{\gamma\delta} \omega_\gamma^\mu\omega_\delta^\mu\Bigg)\rme^{-\rmi \frac{\kappa}{\sigma}\sum_{\gamma=1}^n\omega_{\gamma}^{\mu}}\prod_{\gamma=1}^n \hat{f}_{y_\gamma^\mu}(\omega_{\gamma}^\mu),
	\label{eq:aveQuant1}
\end{align}
where $q_{\gamma\delta}$ are the order parameters introduced in~\eqref{eq:qdef} 

Using equation \eqref{eq:fourier} we get:
\begin{align}\notag
	&\left\langle \prod_{\gamma=1}^n \exp\left(-\rmi y_\gamma^\mu \Phi\left(\frac{\wv^\gamma\cdot\xv^\mu}{\sigma \sqrt{N}} -\frac{\kappa}{\sigma}\right)\right)\right\rangle_{\Pi}\\
	\notag
	&\quad\simeq\frac{1}{(2\pi)^{n}}\int_{\mathbb{R}^n} \left(\prod_{\gamma=1}^n \de \lambda_{\gamma}^\mu\right) \rme^{-\rmi \sum_{\gamma=1}^n y_\gamma^\mu \Phi(\lambda_\gamma^\mu) }\\
	&\qquad \qquad \times \int_{\mathbb{R}^n} \left(\prod_{\gamma=1}^n \de \omega_{\gamma}^\mu \right) \rme^{-\rmi \sum_{\gamma=1}^n \left(\kappa/\sigma+\lambda_\gamma^\mu\right)\omega_{\gamma}^{\mu}}  \exp\left(-\frac{1}{2\sigma^2 }\sum_{\gamma,\delta=1}^n q_{\gamma\delta} \omega_\gamma^\mu\omega_\delta^\mu\right).
\end{align}
The computation of the second term in square brakets of equation \eqref{eq:aveQuant0} is completely analogous and leads to the same result. We can thus write
\begin{align}\notag
	&\left\langle \prod_{\gamma=1}^n \rme^{-\rmi y_\gamma^\mu R^\mu_\gamma(\kappa)}\right\rangle_{\Pi,\xiv}\simeq \int_{\mathbb{R}^n} \left(\prod_{\gamma=1}^n \frac{\de \lambda_{\gamma}^\mu}{2\pi}\right)\int_{\mathbb{R}^n}\left(\prod_{\gamma=1}^n \de \omega_{\gamma}^\mu \right)  \\
	&\qquad \qquad \times  \exp\left(-\rmi \sum_{\gamma=1}^n y_\gamma^\mu \Phi(\lambda_\gamma^\mu) -\rmi \sum_{\gamma=1}^n \left(\frac{\kappa}{\sigma}+\lambda_\gamma^\mu\right)\omega_{\gamma}^{\mu}-\frac{1}{2\sigma^2 }\sum_{\gamma,\delta=1}^n q_{\gamma\delta} \omega_\gamma^\mu\omega_\delta^\mu\right)
	\label{eq:aveQuantumTheta}
\end{align}
Which can be inserted into \eqref{eq:avethetaInt} to finally obtain:
\begin{align}\notag
	&\left\langle \prod_{\mu=1}^p\prod_{\gamma=1}^n \theta \left(R^\mu_\gamma(\kappa) -1+\epsilon\right)\right\rangle \\
	\notag
	&\qquad \simeq
	\prod_{\mu=1}^p \int_{[1-\epsilon,\infty)^n} \left(\prod_{\gamma=1}^n\frac{\de z_\gamma^\mu}{2\pi}\right)\int_{\mathbb{R}^n\times\mathbb{R}^n\times \mathbb{R}^n} \left(\prod_{\gamma=1}^n \frac{\de \lambda_{\gamma}^\mu \de y_\gamma^\mu \de \omega_{\gamma}^\mu}{2\pi}\right)\rme^{K(\{\lambda_\gamma^\mu\},\{y_\gamma^\mu\},\{\omega_\gamma^\mu\},\{q_{\gamma\delta}\})},
	\label{eq:avethetaFinalapp}
\end{align}
where $K$ is given in \eqref{eq:K}. Then, using the fact that the factors in the product corresponding to each $\mu=1,\dots,p$ are all equal, we can drop the $\mu$ index from the integration variables and finally obtain~\eqref{eq:avethetaFinal}.

\section{Derivation of $G_1^{\RS}(q)$}
\label{app:G1RS}

The restriction of \eqref{eq:K} to the replica symmetric subspace gives
\begin{equation}\label{eq:KRS}
	K^{RS}(\{\lambda_\gamma\},\{y_\gamma\},\{\omega_\gamma\},q)=\rmi \sum_{\gamma=1}^n y_\gamma \left[z_\gamma- \Phi(\lambda_\gamma)\right] -\rmi \sum_{\gamma=1}^n \left(\frac{\kappa}{\sigma}+\lambda_\gamma\right)\omega_{\gamma}-\frac{1-q}{2\sigma^2}\sum_{\gamma=1}^n \omega_\gamma^2-\frac{q}{2\sigma^2 }\Bigg(\sum_{\gamma=1}^n  \omega_\gamma\Bigg)^2
\end{equation}
and the integration over the $\omega_\gamma$ variables in \eqref{eq:G10} can now be carried out, yielding:
\begin{align}\notag
	&\int_{\mathbb{R}^n} \left(\prod_{\gamma=1}^n \de \omega_{\gamma}  \right) \exp\Bigg(-\rmi \sum_{\mu=1}^p \sum_{\gamma=1}^n \left(\frac{\kappa}{\sigma}+\lambda_\gamma \right)\omega_{\gamma}^{\mu}-\frac{q}{2\sigma^2 }\sum_{\substack{\gamma,\delta=1\\ \gamma\neq \delta}}^n\  \omega_\gamma \omega_\delta -\frac{1}{2\sigma^2}\sum_{\gamma=1}^n (\omega_\gamma )^2\Bigg)\\
	\notag
	&\quad=\sigma^n \int_{\mathbb{R}^n}\Bigg(\prod_{\gamma=1}^n\de \widetilde{\omega}_{\gamma}  \Bigg) \exp\Bigg(-\rmi \sum_{\gamma=1}^n\left(\kappa+\sigma \lambda_\gamma \right)\widetilde{\omega}_{\gamma}^{\mu}-\frac{q}{2}\Bigg(\sum_{\gamma=1}^n \widetilde{\omega}_{\gamma}  \Bigg)^2-\frac{1-q}{2}\sum_{\gamma=1}^n(\widetilde{\omega}_{\gamma} )^2\Bigg) \\
	&\quad= \int_{-\infty}^{+\infty} \de t\frac{\rme^{-t^2/2}}{\sqrt{2\pi}}\prod_{\gamma=1}^n \left[ \sqrt{\frac{2\pi \sigma^2}{1-q}}\exp\left(-\frac{(\kappa+\lambda_\gamma \sigma + \sqrt{q} t)^2}{2(1-q)}\right)\right],
	\label{eq:OmegaIntegration}
\end{align}
where the first equality follows from a change of variables and the second one follows from a gaussian integration after a linearization of the quadratic term using the Gaussian integral trick 
\begin{equation}\label{eq:gaussianIntTrick}
	\rme^{-\frac{q}{2}\left(\sum_{\gamma=1}^n \omega_\gamma\right)^2}=\int_{-\infty}^{\infty} \frac{\de t}{\sqrt{2\pi}} \rme^{-\frac{t^2}{2}+\rmi \frac{\sqrt{q}}t\sum_{\gamma=1}^n \omega_\gamma}.
\end{equation}
Using \eqref{eq:OmegaIntegration} into \eqref{eq:G10} and a rescaling of the $\lambda$ variables, we can write the integral inside the logarithm in \eqref{eq:G10} as:
\begin{align}\notag
	&\int_{-\infty}^{+\infty} \de t \frac{\rme^{-t^2/2}}{\sqrt{2\pi}}
	\prod_{\gamma=1}^n\int_{-\infty}^{+\infty} \frac{\de \lambda_\gamma}{\sqrt{2\pi (1-q)}}\int_{1-\epsilon}^\infty \frac{\de  z_\gamma}{2\pi}\int \de y_\gamma~ \rme^{-\rmi y_\gamma[ z_\gamma -\Phi(\lambda_\gamma/\sigma)]}\exp\left(-\frac{(\kappa+\lambda_\gamma+\sqrt{q}t)^2}{2(1-q)}\right)\\
	\notag
	&=\int_{-\infty}^{+\infty} \de t \frac{\rme^{-t^2/2}}{\sqrt{2\pi}}\left(\int_{-\infty}^{+\infty} \frac{\de \lambda}{\sqrt{2\pi (1-q)}}\int_{1-\epsilon}^\infty \frac{\de  z}{2\pi}\int \de y~ \rme^{-\rmi y[ z -\Phi(\lambda/\sigma)]}\exp\left(-\frac{(\kappa+\lambda+\sqrt{q}t)^2}{2(1-q)}\right)\right)^n\\
	&=\int_{-\infty}^{+\infty} \de t \frac{\rme^{-t^2/2}}{\sqrt{2\pi}}\left(\int_{-\infty}^{+\infty} \frac{\de \lambda}{\sqrt{2\pi (1-q)}}\theta\left[\Phi\left(\frac{\lambda}{\sigma}\right)-1+\epsilon\right]\exp\left(-\frac{(\kappa+\lambda+\sqrt{q}t)^2}{2(1-q)}\right)\right)^n
	\label{eq:INT}
\end{align}
where the first equality follows from the fact that factors corresponding to different $\gamma$'s are all equal to each other, and the second equality follows from the integral representation of the $\theta$ function. After a further change of the integration variables, \eqref{eq:INT} can be rewritten as
\begin{align}\notag
	&\int_{-\infty}^{+\infty} \de t \frac{\rme^{-t^2/2}}{\sqrt{2\pi}}\left(\int_{\kappa+\sigma\Phi^{-1}(1-\epsilon)}^\infty \frac{\de \lambda}{\sqrt{2\pi (1-q)}}\exp\left(-\frac{(\lambda+\sqrt{q}t)^2}{2(1-q)}\right)\right)^n\\
	&\qquad=\int_{-\infty}^{+\infty} \de t \frac{\rme^{-t^2/2}}{\sqrt{2\pi}}\left(1-\Phi\left(\widetilde{\kappa}+\frac{\sqrt{q}t}{\sqrt{1-q}} \right)\right)^n,
\end{align}
which is the argument of the logarithm in \eqref{eq:G1RS}.

\end{document}